\documentclass[12pt]{article}

\usepackage[utf8]{inputenc}
\usepackage{latexsym, graphicx} 
\usepackage{amsmath,amsfonts,amssymb}
\usepackage{slashed}
\usepackage{cancel}
\usepackage{mathrsfs}
\usepackage{tensor}
\usepackage{xcolor}
\usepackage{cite}
\usepackage[colorlinks=true,linkcolor=blue,citecolor=blue,urlcolor=blue,filecolor=black]{hyperref}

\renewenvironment{thebibliography}[1]{
  \begin{oldthebibliography}{#1}
    \setlength{\itemsep}{2pt}
    \setlength{\parskip}{2pt}
}
{
  \end{oldthebibliography}
}

\numberwithin{equation}{section} 

\newcommand*{\scri}{\ensuremath{\mathscr{I}}}

\newcommand*{\dd}{\mathop{}\!d}
\newcommand*{\sgn}{\text{sgn}}

\begin{document}

\begin{titlepage}
  \thispagestyle{empty}

\begin{flushright}
\end{flushright}

\vskip1cm

\begin{center}  
{\Large\textbf{Carrollian conformal correlators\\ \vskip 5mm and massless scattering amplitudes}}

\vskip1cm

\centerline{Kevin Nguyen}

\vskip1cm

{\it{Universit\'e Libre de Bruxelles, ULB-Campus Plaine
CP231, 1050 Brussels, Belgium}}\\
\vskip 1cm
{kevin.nguyen2@ulb.be}

\end{center}

\vskip1cm

\begin{abstract} 
The theory of particle scattering is concerned with transition amplitudes between states that belong to unitary representations of the Poincar\'e group. The latter acts as the isometry group of Minkowski spacetime $\mathbb{M}$, making natural the introduction of relativistic tensor fields encoding the particles of interest. Since the Poincar\'e group also acts as a group of conformal isometries of null infinity~$\scri$, massless particles can also be very naturally encoded into Carrollian conformal fields living on $\scri$. In this work we classify the two- and three-point correlation functions such Carrollian conformal fields can have in any consistent quantum theory of massless particles and arbitrary dimension. We then show that bulk correlators of massless fields in~$\mathbb{M}$ explicitly reduce to these Carrollian conformal correlators when evaluated on $\scri$, although in the case of time-ordered bulk correlators this procedure appears singular at first sight. However we show that the Carrollian correlators of the descendant fields are perfectly regular and precisely carry the information about the corresponding S-matrix elements.
\end{abstract}

\end{titlepage}

{\hypersetup{linkcolor=black}
  \tableofcontents 
}

\section{Introduction}
Scattering theory has recently been the focus of renewed interest initially triggered by the observation that soft graviton theorems are Ward identities \cite{Strominger:2013jfa,He:2014laa,Adamo:2014yya,Kapec:2014opa,Campiglia:2014yka,Campiglia:2015kxa,Donnay:2022hkf,Agrawal:2023zea} associated with the asymptotic symmetries of General Relativity with flat asymptotics \cite{Bondi:1962px,Sachs:1962wk,Sachs:1962zza,Barnich:2009se,Barnich:2010eb}. In particular the subleading soft graviton theorem derived in \cite{Cachazo:2014fwa} was recast in the form of the conformal Ward identity of two-dimensional conformal field theory (CFT) \cite{Kapec:2016jld}. Given the power of conformal symmetry, the question arose whether this could be exploited in the study of gravitational scattering amplitudes and quantum gravity more broadly, a topic now referred to as \textit{celestial holography}. For recent reviews of this subject we refer to \cite{Pasterski:2021raf,McLoughlin:2022ljp,Donnay:2023mrd}. One of the key ideas of celestial holography is to represent scattering amplitudes in a basis of boost rather than momentum eigenstates, such that they manifestly take the form of correlation functions in a somewhat exotic two-dimensional CFT \cite{deBoer:2003vf,Pasterski:2016qvg,Pasterski:2017kqt,Pasterski:2017ylz}. It is however still unclear whether scattering amplitudes can be fully formulated in the language of two-dimensional CFT, and if so, what are the basic axioms satisfied by such a theory. 

Motivated by these observations we may want to take a second look at the Poincar\'e group underlying the scattering theory of particles. Its Lorentz subgroup $SO(1,3)=SL(2,\mathbb{C})$ is also the group of global conformal transformations of the (celestial) sphere $S^2$, and the approach pursued in celestial holography is precisely to organise the scattering amplitudes accordingly. However this happens at the cost of obscuring the action of translations, as the latter end up acting as internal symmetries connecting conformal primaries of different scaling dimensions. To avoid this drawback one can instead choose to realise the full Poincar\'e group as the conformal group of a three-dimensional albeit degenerate manifold $\scri \simeq \mathbb{R} \times S^2$ that can be identified with the null conformal boundary of Minkowski spacetime. This allows to recast \textit{massless} scattering amplitudes as a set of correlators of a three-dimensional conformal theory living on $\scri$, an alternative approach called \textit{Carrollian holography} \cite{Bagchi:2016bcd,Banerjee:2018gce,Banerjee:2019prz,Bagchi:2022emh,Donnay:2022aba,Donnay:2022wvx,Saha:2023hsl,Saha:2023abr,Bagchi:2023cen}. At this point it is however equally unclear whether massless scattering amplitudes can be fully formulated using the  language of Carrollian conformal field theory, especially given that the latter has not yet been given a general formulation. The present work is part of an effort to construct the general framework of Carrollian CFT, which will hopefully provide us with an independent `holographic' formulation of scattering theory.

We therefore aim to revisit the standard theory of massless particle scattering through Carrollian conformal field theory rather than relativistic quantum field theory. At the most basic level we are still concerned with quantum systems whose Hilbert space consists of unitary irreducible representations (UIR) of the Poincar\'e group $ISO(1,3)$ \cite{Wigner:1939cj}. In this context quantum fields are usually introduced as a mean to conveniently handle locality and Lorentz covariance at the cost of manifest unitarity, especially with regards to the construction of interactions consistent with Poincar\'e invariance and the cluster decomposition principle \cite{Weinberg:1995mt}. Even though they are extremely useful quantities, relativistic quantum fields are however not fundamental by contrast to the particles they encode. The definition of the S-matrix from which physical observables are constructed in fact does not even require the introduction of quantum fields. Hence their introduction is a matter of taste and convenience. We are  free to consider a new kind of quantum fields encoding the very same massless particles, namely the Carrollian conformal fields living on $\scri$ \cite{Banerjee:2018gce,Donnay:2022wvx,Nguyen:2023vfz}.

First we need to develop the general framework of Carrollian CFT, and later on study how scattering amplitudes fit into this framework. As mentioned earlier we will concern ourselves with quantum systems whose Hilbert space consists of massless UIRs of the Poincar\'e group $ISO(1,d)$ for arbitrary dimension $d$ that can be encoded into Carrollian conformal fields at $\scri\simeq \mathbb{R} \times S^{d-1}$ as previously shown in \cite{Nguyen:2023vfz}. Proceeding by analogy with standard conformal field theory, we then want to classify the two- and three-point correlation functions allowed by Poincar\'e symmetry. For $d=3$ the two-point functions were classified in \cite{Donnay:2022wvx,Salzer:2023jqv} while an incomplete set of three-point functions has been described in \cite{Salzer:2023jqv,Bagchi:2023fbj}. See also \cite{Banerjee:2018gce,Chen:2021xkw} for earlier discussions. Armed with these basic ingredients we may then want to determine the operator product expansion (OPE), conformal block decomposition, crossing equations, et cetera, that should severely constrain the theory and therefore provide us with predictive power. In this work we will however restrict to the classification of two- and three-point functions. Once this is done, we will look for specific theories that can provide us with explicit realisations of these Carrollian correlators. Since this was the main motivation for this work, we could directly look at scattering amplitudes themselves as in \cite{Banerjee:2018gce,Banerjee:2019prz,Bagchi:2022emh,Donnay:2022wvx}. There currently exists two versions of the transformation mapping scattering amplitudes to Carrollian correlators. The first one, introduced as \textit{modified Mellin transform} in \cite{Banerjee:2018gce}, maps massless particle UIRs to Carrollian conformal fields of arbitrary scaling dimension. The second one corresponds to fixing the scaling dimension to $\Delta=1$ (for $d=3$), in which case the modified Mellin transform reduces to a simple Fourier transform with respect to the energy of the particles  \cite{Donnay:2022wvx}. The reason for this particular choice of scaling dimension is that massless bulk fields in $\mathbb{M}^{d+1}$ reduce to Carrollian conformal fields with $\Delta=\frac{d-1}{2}$ when pulled back to $\scri$ \cite{Donnay:2022wvx,Nguyen:2023vfz}. However this specific choice of scaling dimension already runs into trouble for the trivial scattering, namely that of a free particle propagating with unit probability. Indeed the integral transform turns out to be divergent and regulators need to be introduced \cite{Donnay:2022wvx}. Furthermore the S-matrix element for three massless particles vanishes as a distribution as it only receives contribution from the colinear limit. Hence both two- and three-point Carrollian functions defined in this way appear very delicate. To gain some insight into these problematic features, we will make a detour via the correlators\footnote{Note that similar issues in celestial holography are avoided by looking at time-ordered correlators rather than scattering amplitudes \cite{Sleight:2023ojm}.} of massless fields in Minkowski spacetime $\mathbb{M}^{d+1}$ and evaluate them in the limit where all insertion points are sent out to null infinity $\scri$. Doing this we will generically recover the correlators of the Carrollian conformal fields encoding the corresponding massless particles, a result which appears very natural given the established correspondence between massless bulk fields and Carrollian conformal fields \cite{Donnay:2022wvx,Nguyen:2023vfz}. However in the case of time-ordered correlators, which are closely related to scattering amplitudes through the LSZ reduction formula \cite{LSZ}, we will see re-emerge the same problematic features as mentioned above. In particular we will observe the appearance of contact terms with anomalous $\ln r$ asymptotic behavior.\footnote{Similar observations have been made in \cite{Donnay:2022ijr,Kim:2023qbl}.} We find that these can be eliminated by acting with $\partial_u$ derivatives, i.e., by considering correlators of \textit{descendant} Carrollian conformal fields. Doing this we will be left with perfectly regular Carrollian two- and three-point functions, with the latter only receiving contributions from the colinear limit. We will finally show that these results are in agreement with those obtained from the corresponding S-matrix elements through the integral transform proposed in \cite{Banerjee:2018gce,Bagchi:2022emh} as long as we restrict our attention to these descendant fields.

The paper is organised in the following way. In section~\ref{section 2} we start by reviewing the construction of the Carrollian conformal fields as field representations of $ISO(1,d)$ living on $\scri \simeq \mathbb{R} \times S^{d-1}$, and the way they encode massless particle UIRs \cite{Nguyen:2023vfz}. We then consider the Ward identities imposed by Poincar\'e invariance on the correlation functions such Carrollian conformal fields can have in any consistent quantum theory. We classify the two-point solutions for arbitrary spin and the three-point solutions for zero spin. Their form is not uniquely fixed by symmetry as was already pointed out in \cite{Bagchi:2022emh,Donnay:2022wvx}. In section~\ref{section 3} we consider examples of two- and three-point correlators of massless fields in $\mathbb{M}^{d+1}$ and evaluate them in the limit where all insertion points are sent out to null infinity $\scri$. More specifically we consider the Wightman, retarded and Feynman propagators associated with scalar, photon and graviton fields in any spacetime dimension, as well as the time-ordered three-point function of a massless scalar with cubic interaction in dimension $d=5$. In all cases evaluation at $\scri$ yields the Carrollian correlators studied in section~\ref{section 2}, provided we act with $\partial_u$ derivatives in case contact singularities appear. We observe a distinction between even and odd $d$ related to the violation of Huygens principle in odd spacetime dimension \cite{Balazs,Soodak1993WakesAW}. We also observe that the free coefficients of the Carrollian two-point functions are determined by the choice of causality of the bulk correlator (Wightman, retarded, Feynman). Finally in section~\ref{section 4} we discuss the relation to scattering amplitudes. In particular we observe that the issue of divergent integrals arising when turning S-matrix elements into Carrollian correlators \cite{Donnay:2022wvx} actually disappears when considering descendant fields, and yields Carrollian correlators that agree with those obtained in section~\ref{section 3}. This provides further evidence that massless scattering theory can be framed within Carrollian conformal field theory.

\section{Carrollian conformal Ward identities} \label{section 2}
The Carrollian conformal fields of interest in this work are those encoding the massless particle UIRs of the Poincar\'e group $ISO(1,d)$. We start by giving a brief account of these fields and refer to \cite{Nguyen:2023vfz} for a complete discussion. The Poincar\'e algebra $\mathfrak{iso}(1,d)$ is usually expressed in the basis of generators $\{\tilde P_\mu\,, \tilde J_{\mu\nu}\}$ which acquire the meaning of generators of spacetime translations and Lorentz transformations. Indeed Minkowski spacetime $\mathbb{M}^{d+1}$ is realised as the homogeneous space
\begin{equation}
\mathbb{M}^{d+1}=\frac{ISO(1,d)}{SO(1,d)}\,.
\end{equation}
on which Poincar\'e transformations act as isometries. 
To discuss Carrollian conformal fields it is however more natural to discuss the homogeneous space \cite{Herfray:2020rvq,Figueroa-OFarrill:2021sxz}
\begin{equation}
\scri=\frac{ISO(1,d)}{(ISO(d-1) \ltimes \mathbb{R}^d) \rtimes \mathbb{R}}\,,
\end{equation}
on which Poincar\'e symmetries are rather realised as conformal isometries. The homogeneous space $\scri$ has the topology $\mathbb{R} \times S^{d-1}$ and can further be identified with the null conformal boundary of $\mathbb{M}^{d+1}$. See \cite{Nguyen:2022zgs} and references therein for a geometrical description of $\scri$ in the context of asymptotically flat gravity. The subgroups $ISO(d-1)\,, \mathbb{R}^d$ and $\mathbb{R}$ are respectively generated by $\{J_{ij},B_i\}$,  $\{K_i,K\}$ and $\{D\}$, defined in terms of the standard basis $\{ \tilde J_{\mu\nu}, \tilde P_\mu\}$ through
\begin{equation}
\label{tilde J}
\tilde J_{ij}=J_{ij}\,, \qquad \tilde J_{i0}=-\frac{1}{2}\left(P_i+K_i \right)\,, \qquad \tilde J_{id}= \frac{1}{2}\left(P_i-K_i \right)\,, \qquad \tilde J_{0d}=-D\,,
\end{equation}
and
\begin{equation}
\label{tilde P}
\tilde P_0=\frac{1}{\sqrt{2}}(H+K)\,, \qquad \tilde P_i=-\sqrt{2}\, B_i\,, \qquad \tilde P_d=\frac{1}{\sqrt{2}}(K-H)\,.
\end{equation}
The commutation relations between these elements of the Poincar\'e algebra may be found in \cite{Nguyen:2023vfz}. 

We are interested in field representations of the Poincar\'e group living on $\scri$, which we call Carrollian conformal fields in reference to the null and conformal nature of $\scri$. A finite-component Carrollian conformal field representation of $ISO(1,d)$ is fully determined from a spin-$s$ tensor representation $\Sigma$ of $SO(d-1)$ and a scaling dimension $\Delta \in \mathbb{R}$,
\begin{equation}
\label{inducing rep}
\left[J_{ij}, \phi_{i_1...i_s}\right]=(\Sigma_{ij})_{i_1...i_s}{}^{\, j_1...j_s}\, \phi_{j_1...j_s}\,, \qquad \left[D,\phi_{i_1...i_s} \right]=i\Delta\, \phi_{i_1...i_s}\,. 
\end{equation}
In case $\Sigma$ is a tensor-spin representation, one should add a spinor index to these equations.
Then the full set of infinitesimal Poincar\'e transformations on a Carrollian conformal field $\phi(x)\equiv \phi_{i_1...i_s}(x^\alpha)$ defined over $\scri$ is given by
\begin{equation}
\label{Carrollian induced rep}
\begin{split}
\left[H, \phi(x)\right]&=i \partial_u \phi(x)\,,\\
\left[P_i, \phi(x)\right]&=i \partial_i \phi(x)\,,\\
\left[J_{ij}, \phi(x)\right]&=i\left(-x_i \partial_j+x_j\partial_i-i\Sigma_{ij} \right)\phi(x)\,,\\
\left[D, \phi(x)\right]&=i\left(\Delta+u\partial_u+x^i\partial_i \right)\phi(x)\,,\\
\left[K, \phi(x)\right]&=i x^2\partial_u \phi(x)\,,\\
\left[K_i, \phi(x)\right]&=i\left(-2x_i \Delta-2ix^j \Sigma_{ij}-2x_i u\partial_u-2x_i x^j \partial_j+x^2 \partial_i \right)\phi(x)\,,\\
\left[B_i, \phi(x)\right]&=ix_i\partial_u \phi(x)\,.
\end{split}
\end{equation}
The defining relations \eqref{inducing rep} correspond to the transformations at the origin $x^\alpha=0$. Let us now describe the connection between these Carrollian conformal fields and the massless particle UIRs \cite{Nguyen:2023vfz}. Following Wigner's foundational work \cite{Wigner:1939cj}, to construct the particle states one first chooses a reference momentum vector $p^{(0)}_\mu=(-1,0_i,1)/\sqrt{2}$ and then picks an inducing irreducible tensor-spin representation of the short little group $SO(d-1)$,
\begin{equation}
\psi_\sigma(p^{(0)})\equiv |p^{(0)},\sigma\rangle\,,
\end{equation}
where $\sigma=i_1\, ...\, i_s$ labels the tensor-spin components of the representation. The full induced representation is then obtained by `boosting' these states to give them arbitrary null momentum,
\begin{equation}
\label{psi(p)}
\psi_\sigma(p)\equiv e^{-ix^i P_i}\, e^{i\ln \omega D}\, \psi_\sigma(p^{(0)})\,.
\end{equation}
The generators $P_i$  and $D$ are precisely the boost generators that do not leave the reference momentum $p^{(0)}$ invariant as can be seen using \eqref{tilde J}. Here $(x^i,\omega)$ parametrise the null momentum of this generic state,
\begin{equation}
\label{momentum parametrisation}
p^\mu(\omega,x^i)=\frac{\omega}{\sqrt{2}}(1+x^2,2x^i,1-x^2)\equiv \omega\, \hat q^\mu(x^i)\,,
\end{equation}
as can be explicitly derived by acting with $\tilde P^\mu$ on \eqref{psi(p)} as shown in \cite{Nguyen:2023vfz,Iacobacci:2024laa}. To connect with the Carrollian conformal fields we need to perform the following integral transform \cite{Banerjee:2018gce,Nguyen:2023vfz},
\begin{equation}
\label{Fourier transform}
\psi^\Delta_\sigma(u,x^i)\equiv \int_0^\infty d\omega\, \omega^{\Delta-1} e^{i\omega u}\, \psi_\sigma(\omega,x^i)\,.
\end{equation}
The above states transform precisely like the Carrollian conformal fields, and we therefore have the state-operator correspondence\footnote{Here we do not claim that any state can be represented by a local operator insertion on $\scri$. Proving or disproving this constitutes an open problem of prime importance.}
\begin{equation}
\label{state operator}
\phi^\Delta_\sigma(x^\alpha)|0\rangle=\psi^\Delta_\sigma(x^\alpha)\,.
\end{equation}
The situation described here should be contrasted with that of massless relativistic fields in $\mathbb{M}^{d+1}$ that are usually used to encode the massless particle states. In the latter case the encoding is intrincate as it involves the embedding of a UIR into a reducible and non-unitary $SO(1,d)$ tensor-spin representation, and the subsequent elimination of the superfluous degrees of freedom through transversality conditions and gauge redundancies while imposing the massless wave equation. Nothing of this sort is required for encoding massless particles into Carrollian conformal fields as the number of degrees of freedom perfectly match, allowing in particular for the relation \eqref{state operator} to hold trivially. There is also no need for a wave equation or any Carrollian analogue as the quadratic Casimir $\mathcal{C}_2=\tilde P_\mu \tilde P^\mu$ automatically vanishes for these Carrollian field representations \cite{Nguyen:2023vfz}. 

Up to this point the choice of scaling dimension $\Delta$ was left totally unconstrained. As we will review at the beginning of section~\ref{section 3}, starting from relativistic massless fields in $\mathbb{M}^{d+1}$ and pulling them back to $\scri$ viewed as its null conformal boundary, one precisely recovers \eqref{Fourier transform} with the particular value
\begin{equation}
\label{special delta}
\Delta=\frac{d-1}{2}\,.
\end{equation}
This value of the scaling dimension will be the preferred one when connecting the Carrollian framework to Minkowskian physics, although other choices of $\Delta$ may also find useful applications. Also note that the Carroll translations $P_\alpha=(H,P_i)$ are raising operators for the scaling dimension $\Delta$, while special conformal transformations $K_\alpha=(K,K_i)$ are lowering operators \cite{Nguyen:2023vfz}. In particular time translation simply acts by 
\begin{equation}
H\, \psi^\Delta_\sigma(x^\alpha)=i\partial_u \psi_\sigma^\Delta(x^\alpha)=-\int_0^\infty d\omega\, \omega^{\Delta}\, e^{i\omega u}\, \psi_\sigma(\omega,x^i)=-\psi^{\Delta+1}_\sigma(x^\alpha)\,,
\end{equation}
or in terms of the corresponding field,
\begin{equation}
\label{descendant}
[H,\phi^\Delta_\sigma(x^\alpha)]=i\partial_u \phi^\Delta_\sigma(x^\alpha)=-\phi^{\Delta+1}_\sigma(x^\alpha)\,.
\end{equation}
We call $\phi^{\Delta+1}_\sigma(x^\alpha)$ a \textit{descendant field} like in standard conformal field theory.

Invariance under $ISO(1,d)$ imposes kinematic constraints on correlations functions such Carrollian conformal fields can have in any consistent quantum theory. More specifically, any $n$-point correlator must satisfy the integrated Ward identity
\begin{equation}
\label{Ward identities}
\sum_{i=1}^n \langle \phi_1(x_1)\,...\,\delta\phi_i(x_i)\,...\, \phi_n(x_n) \rangle=0\,,
\end{equation}
where $\delta \phi_i$ is any linear combination of the infinitesimal transformations \eqref{Carrollian induced rep}. In particular Carroll translation invariance generated by $P_\alpha=(H,P_i)$ implies that correlation functions only depend on the insertion points through $x^\alpha_{ij}\equiv x^\alpha_i-x^\alpha_j$. As a set of indepedent variables we can therefore choose $x^\alpha_i-x^\alpha_{i+1}$ with $i=1,...,n-1$. We will now classify the two- and three-point solutions to the Ward identities \eqref{Ward identities}.  

\subsection{Spinning two-point solutions}
We start by analysing the constraints imposed on generic two-point functions
\begin{equation}
C_{\sigma_1 \sigma_2}^{(2)}(x_{12})\equiv \langle \phi_{\sigma_1}(x_1)\phi_{\sigma_2}(x_2) \rangle\,.
\end{equation}
A complete analysis for $d=3$ was already presented in \cite{Donnay:2022wvx}, which we now generalise to arbitrary dimension. The Ward identity \eqref{Ward identities} associated with the Carroll boost $B_i$ can be written
\begin{equation}
x_{12}^i\, \partial_{u_{12}} C_{\sigma_1 \sigma_2}^{(2)}=0\,,
\end{equation}
whose distributional solutions are simply
\begin{equation}
\label{C2 solution}
C_{\sigma_1 \sigma_2}^{(2)}(x_{12}^\alpha)=f_{\sigma_1 \sigma_2}(x_{12}^i)+  g_{\sigma_1 \sigma_2}(u_{12})\, \delta(x_{12}^i)\,.
\end{equation}
Without loss of generality we can assume that all contact terms are captured by the second term, such that $f_{\sigma_1 \sigma_2}(x^i)$ should independently solve the Ward identities \eqref{Ward identities}. Looking at \eqref{Carrollian induced rep} we see that for this term they reduce to the standard Ward identities of conformal field theory in $\mathbb{R}^{d-1}$, with well-known solutions\begin{equation}
\label{CFT 2-point function}
f_{\sigma_1 \sigma_2}(x)=\delta_{\Delta_1,\Delta_2}\, \delta_{\Sigma_1,-\Sigma_2}\, \frac{\mathcal{I}_{\sigma_1 \sigma_2}(x)}{|x|^{\Delta_1+\Delta_2}}\,.
\end{equation}
In particular a non-zero solution requires the tensor-spin representations of the two fields to be conjugate of each other, namely $\Sigma_{ij}^1=-\Sigma_{ij}^2$, and their scaling dimensions to be identical. The explicit expression for the inversion tensors $\mathcal{I}_{\sigma_1 \sigma_2}(x)$ may be found for example in \cite{Osborn:1993cr}, which reduces at $x=0$ to the identity operator $\delta_{\sigma_1 \sigma_2}$ within the subspace of independent components of the $SO(d-1)$ tensor-spin representation. In the case of a totally symmetric and traceless spin-$s$ tensor field $\phi_{i_1\,...\,i_s}$ for example, the identity operator $\delta_{\sigma_1 \sigma_2}$ is explicitly given by the projector onto totally symmetric and traceless tensors. The fact that the solutions \eqref{CFT 2-point function} are allowed is unsurprising given that the Carrollian conformal fields carry representations of the Lorentz subgroup $SO(1,d)$ that coincides with the euclidean conformal group in $d-1$ dimensions. 

We then turn to the determination of the other branch of solutions given in terms of $g_{\sigma_1 \sigma_2}(u)$. Rotation invariance again imposes the tensor-spin representations of the two fields to be conjugate of each other, and we can therefore write
\begin{equation}
\label{C2 solution bis}
C_{\sigma_1\sigma_2}^{(2)}(u,x^i)=
f_{\sigma_1\sigma_2}(x^i)+\delta_{\sigma_1\sigma_2}\, g(u)\, \delta(x^i)\,.
\end{equation}
Applied to this set of solutions the dilation Ward identity then implies
\begin{equation}
\label{dilation Ward id}
\left[\Delta_1+\Delta_2+u\partial_u+x^i \partial_i\right]\delta(x^i)\, g(u)=0\,,
\end{equation}
which reduces to 
\begin{equation}
\label{dilation Ward id 2}
\left[\Delta_1+\Delta_2+u\partial_u-(d-1)\right]g(u)=0\,,
\end{equation}
after using
\begin{equation}
x^i\partial_i \delta(x^i)=-\partial_i x^i \delta(x^i)=-(d-1)\, \delta(x^i)\,.
\end{equation}
Equation \eqref{dilation Ward id 2} admits the solutions\footnote{Note that the solution $1/u^{\Delta_1+\Delta_2+1-d}$ needs to be properly regulated at $u=0$. At the moment we do not see a clear principle to help us choose a regularisation scheme so we will leave it undetermined.}
\begin{equation}
\label{g(u) solutions}
g(u)=\begin{cases}
a+b\, \sgn(u) & \text{if } \Delta_1+\Delta_2=d-1\\
\frac{a}{u^{\Delta_1+\Delta_2+1-d}} &  \text{if } \Delta_1+\Delta_2\neq d-1 \,.
\end{cases}
\end{equation}
It turns out that the case $\Delta_1+\Delta_2=d-1$ will precisely be the one of interest when making contact with massless fields in Minkowski spacetime, as can already be seen from \eqref{special delta}. Finally we can check that invariance under special conformal transformation generated by $K_\alpha=(K,K_i)$ do not impose extra restrictions. Indeed $K$-invariance requires
\begin{equation}
\label{K constraint}
(|x_1|^2-|x_2|^2)\, \delta(x_{12}^i)\, g'(u_{12})=0\,,
\end{equation}
which is obviosuly automatically satisfied. Similarly invariance associated with the generators $K_i$ is satisfied thanks to $\Sigma_{ij}^1=-\Sigma_{ij}^2$ together with \eqref{dilation Ward id}. 

Let us summarise the situation. The two-point solutions to the Carrollian conformal Ward identities are given by \eqref{C2 solution bis} together with \eqref{CFT 2-point function} and \eqref{g(u) solutions}. In addition to the correlator $f_{\sigma_1 \sigma_2}(x^i)$ familiar from euclidean conformal field theory in $\mathbb{R}^{d-1}$, one can get up to two additional contact term solutions depending on the value of the scaling dimensions.

\subsection{Scalar three-point solutions}
In this section we classify the scalar three-point functions allowed by Poincar\'e symmetry, and denote them
\begin{equation}
C^{(3)}(x_{12}^\alpha,x_{23}^\alpha)\equiv \langle \phi_1(x_1^\alpha) \phi_2(x_2^\alpha) \phi_3(x_3^\alpha)\rangle\,.
\end{equation}
The classification of three-point functions of generic spinning fields is left to future detailed study. A broad class of solutions to the Ward identities resulting from rotation and Carroll boost invariance, namely
\begin{align}
\left(x_{12}^i\, \partial_{u_{12}}+x_{23}^i\, \partial_{u_{23}} \right)C^{(3)}&=0\,,\\
\left(x_{12}^i\, \partial_{x_{12}^j}-x_{12}^j\, \partial_{x_{12}^i}+x_{23}^i\, \partial_{x_{23}^j}-x_{23}^j\, \partial_{x_{23}^i}\right)C^{(3)}&=0\,,
\end{align}
are products of scalar two-point functions with yet unconstrained scaling parameters,
\begin{equation}
\label{C3 product}
C^{(3)}=C_a^{(2)}(x_{12}^\alpha)\, C_b^{(2)}(x_{23}^\alpha)\, C_c^{(2)}(x_{13}^\alpha)\,,
\end{equation}
i.e.~with
\begin{equation}
\label{C2 cases}
C^{(2)}_a(x^\alpha)=
\begin{cases}
c_1 |x|^{-a}+(c_2+c_3\, \sgn(u))\, \delta(x^i) & \text{if } a=d-1\\
c_1 |x|^{-a}+c_2\, u^{d-1-a}\, \delta(x^i) & \text{if } a\neq d-1 \,.
\end{cases}
\end{equation}
Plugging \eqref{C3 product} into the dilation Ward identity,
\begin{equation}
\left(\Delta_1+\Delta_2+\Delta_3+u_{12}\, \partial_{u_{12}}+u_{23}\, \partial_{u_{23}}+x_{12}^i\, \partial_{x_{12}^i}+x_{23}^i\, \partial_{x_{23}^i} \right) C^{(3)}=0\,,
\end{equation}
we obtain a constraint on the parameters $a,b,c$ familiar from standard conformal field theory,
\begin{equation}
\label{abc}
a+b+c=\Delta_1+\Delta_2+\Delta_3\,.
\end{equation}
We are thus left to impose the constraints stemming from special conformal transformations generated by $K_\alpha=(K,K_i)$. While $K$-invariance is automatically satisfied in a way analogous to \eqref{K constraint}, the constraints associated with $K_i$-invariance will depend on the kind of solutions considered. From \eqref{C3 product}-\eqref{C2 cases} we see that three-point solutions can contain a product of up to three contact terms. We will analyse each case separately.

\paragraph{No contact term.} The solution without contact term takes the form
\begin{equation}
\label{standard 3-point}
C^{(3)}=\frac{1}{|x_{12}|^a|x_{23}|^b|x_{13}|^c}\,.
\end{equation}
Because of its time-independence, the Ward identities \eqref{Ward identities} reduce to that of conformal field theory in $\mathbb{R}^{d-1}$. In particular one can show that special conformal transformations ($K_i$) uniquely determine the parameters $a,b,c$ to be 
\begin{equation}
a=\Delta_1+\Delta_2-\Delta_3\,, \qquad b=\Delta_2+\Delta_3-\Delta_1\,, \qquad c=\Delta_3+\Delta_1-\Delta_2\,,
\end{equation}
thereby automatically satisfying \eqref{abc}.
Hence one possible three-point function is the familiar one from conformal field theory in $\mathbb{R}^{d-1}$.

\paragraph{One contact term.} The generic solutions with one contact term take the form
\begin{equation}
\label{one contact first}
C^{(3)}=\frac{\delta(x_{12}^i)}{u_{12}^{a+1-d}\, |x_{23}|^b |x_{13}|^c}=\frac{\delta(x_{12}^i)}{u_{12}^{a+1-d}\,|x_{23}|^{b+c}}\,,
\end{equation}
or one obtained by some permutation of the indices. Besides the generic solutions \eqref{one contact first}, there are additional ones corresponding the particular choice $a=d-1$, namely
\begin{equation}
\label{one contact second}
C^{(3)}=\sgn(u_{12})\, \frac{\delta(x_{12}^i)}{|x_{23}|^b |x_{13}|^c}=\sgn(u_{12})\, \frac{\delta(x_{12}^i)}{|x_{23}|^{b+c}}\,, \qquad (a=d-1)\,.
\end{equation}
The constraints stemming from the Ward identity of special conformal transformations turn out to be
\begin{equation}
a=\Delta_1+\Delta_2-\Delta_3\,, \qquad b+c=2\Delta_3\,,
\end{equation}
which are just the combinations of parameters needed to characterise these solutions.

\paragraph{Two contact terms.} The solutions with two contact terms contain products such as
\begin{equation}
\frac{\delta(x_{12}^i)\, \delta(x_{23}^i)}{|x_{13}|^c}=\frac{\delta(x_{12}^i)\, \delta(x_{23}^i)}{|x_{12}|^c}\,,
\end{equation}
which are difficult to make sense of for $c\neq 0$. Hence we can restrict to solutions where one parameter is set to zero, say $c=0$, 
\begin{equation}
\label{two contact}
C^{(3)}=\frac{\delta(x_{12}^i)\, \delta(x_{23}^i)}{(u_{12})^{a+1-d}\, (u_{23})^{b+1-d}}\,.
\end{equation}
In that case  $K_i$-invariance requires
\begin{equation}
\label{a and b}
a+b=\Delta_1+\Delta_2+\Delta_3\,,
\end{equation}
which coincides with \eqref{abc} for $c=0$. 
In fact this solution can be further generalised to 
\begin{equation}
\label{two contact bis}
C^{(3)}=\frac{\delta(x_{12}^i)\, \delta(x_{23}^i)}{(u_{12})^{a+1-d}\, (u_{23})^{b+1-d}\, (u_{13})^c}\,,
\end{equation}
subject to
\begin{equation}
a+b+c=\Delta_1+\Delta_2+\Delta_3\,,
\end{equation}
Note that the Carrollian three-point functions \eqref{two contact bis} were presented in \cite{Bagchi:2023fbj} for the particular dimension $d=3$. When in addition the scaling dimensions are such that $a=d-1$ (and similarly for $b$ and $c$) then we again have additional solutions obtained from \eqref{two contact} by the  replacement $1/u_{12}^{a+1-d} \mapsto \sgn(u_{12})$.

\paragraph{Three contact terms.} This last class of solutions takes the simple form
\begin{equation}
\frac{\delta(x_{12}^i)\, \delta(x_{23}^i)\, \delta(x_{13}^i)}{u_{12}^{a+1-d}\, u_{23}^{b+1-d}\, u_{13}^{c+1-d}}\,.
\end{equation}
Again when $a=d-1$ (and similarly for $b,c$) then we again have additional solutions obtained from the  replacement $1/u_{12}^{a+1-d} \mapsto \sgn(u_{12})$.
It can be shown that special conformal invariance only requires \eqref{abc} to hold. 

\section{Holographic Carrollian correlators} \label{section 3}
In the previous section we have classified the spinning two-point functions and the scalar three-point functions satisfying the Ward identites \eqref{Ward identities} resulting from $ISO(1,d)$ invariance of the theory. We now make connection with standard quantum field theory in Minkowski spacetime $\mathbb{M}^{d+1}$. We have shown in previous work that there is a simple correspondence between relativistic massless fields in $\mathbb{M}^{d+1}$ and Carrollian conformal fields at its null conformal boundaries $\scri^\pm \simeq \mathbb{R} \times S^{d-1}$ \cite{Nguyen:2023vfz}. Indeed the asymptotic behavior near $\scri$ of a bulk field $\phi_{\mu_1...\mu_s}$ carrying massless spin-$s$ particles coincides with that of the corresponding Carrollian conformal field $\phi^\Delta_{i_1...i_s}$ with scaling dimension 
\begin{equation}
\label{Delta}
\Delta=\frac{d-1}{2}\,. 
\end{equation}
Let us start by giving a brief overview of this correspondence. To this end we introduce retarded coordinates $(r,u,x^i)$ in $\mathbb{M}^{d+1}$, related to cartesian coordinates $X^\mu$ by
\begin{equation}
\label{coord transf retarded}
X^\mu=u\, n^\mu+r\, \hat q^\mu(x^i)\,,
\end{equation}
where $n^\mu$ and $\hat q^\mu(x^i)$ are null vectors with cartesian components
\begin{equation}
\label{hat q}
\begin{split}
\hat q^\mu(x^i)&= \frac{1}{\sqrt{2}}\left(1+x^2\,, 2x^i\,, 1-x^2\right)\,,\\
n^\mu&=\frac{1}{\sqrt{2}}\left(1,0^i,-1\right)\,.
\end{split}
\end{equation}
The latter satisfy $n \cdot \hat q=-1$ and $\hat q(x) \cdot \hat q(y)=-|x-y|^2$. In retarded coordinates the Minkowski metric takes the simple form
\begin{equation}
\label{flat Bondi gauge}
ds_{\mathbb{M}^{d+1}}^2=\eta_{\mu\nu} \dd X^\mu \dd X^\nu =-2 \dd u \dd r+ 2 r^2 \delta_{ij} \dd x^i \dd x^j\,.
\end{equation}
Past and future null infinities $\scri^\pm$ respectively lie at $r \to \pm \infty$ with degenerate conformal metric 
\begin{equation}
ds^2_{\scri^\pm}=0 \dd u^2+\delta_{ij} \dd x^i \dd x^j\,.
\end{equation}
The set of lines with constant $x^\alpha=(u,x^i)$ coordinates is a congruence of null rays interpolating between $\scri^-$ and $\scri^+$. The interested reader should consult appendix~A of \cite{Donnay:2022wvx} for a detailed comparison with more standard retarded/advanced coordinate systems with round metric representatives on the celestial sphere $S^{d-1}$. Introducing the projector onto the space tangent to the celestial sphere $S^{d-1}$,
\begin{equation}
e^{\mu}_i\equiv \frac{1}{r} \frac{\partial X^\mu}{\partial x^i}=\frac{\partial \hat q^\mu}{\partial x^i}\,,
\end{equation}
and the corresponding components of the massless bulk field\footnote{The field components \eqref{celestial components} differ from the coordinate components $\partial_{i_1} X^{\mu_1}\,...\,\partial_{i_s} X^{\mu_s}\, \phi_{\mu_1...\,\mu_s}$ by a factor of $r^s$. The latter therefore asymptotically transforms like a Carrollian conformal field with shifted scaling dimension $\Delta'=\frac{d-1}{2}-s$ as described in \cite{Nguyen:2023vfz} (with an opposite sign convention). However as we will see the holographic Carrollian correlators are unambiguously those associated with $\Delta=\frac{d-1}{2}$.}
\begin{equation}
\label{celestial components}
\phi_{i_1\,...\,i_s}\equiv e^{\mu_1}_{i_1}...\, e^{\mu_s}_{i_s}\, \phi_{\mu_1\,...\,\mu_s}\,,
\end{equation}
in the asymptotic limit $r \to \pm \infty$ towards $\scri^\pm$ the latter behave as \cite{Nguyen:2023vfz}
\begin{equation}
\label{field asymptotic}
\lim_{r \to \pm \infty} \phi_{i_1...i_s}(r,x^\alpha)=r^{-\Delta}\, \phi^{\Delta,\pm}_{i_1...i_s}(x^\alpha)+\text{subleading}\,,
\end{equation}
with $\Delta$ as given in \eqref{Delta}. It can be shown that $\phi_{i_1...i_s}^{\Delta,\pm}(x^\alpha)$ is indeed a Carrollian conformal field carrying the corresponding massless spin-$s$ particles and therefore satisfying \eqref{state operator}. The components \eqref{celestial components} are the independent and gauge-invariant degrees of freedom of the tensor field $\phi_{\mu_1...\mu_s}$, and they indeed encode all the information about the corresponding massless particles. For obvious reasons $\phi_{i_1...i_s}^{\Delta,\pm}(x^\alpha)$ will be called \textit{in} and \textit{out} Carrollian fields, respectively.

Given the correspondence between relativistic massless fields in $\mathbb{M}^{d+1}$ and Carrollian conformal fields at $\scri$, we should expect that some observables in relativistic quantum field theory can be described in terms of Carrollian conformal correlators. A very natural guess is that the asymptotic limit of bulk correlators takes the form of Carrollian conformal correlators, which is in fact essentially guaranteed by Poincar\'e symmetry. We will demonstrate this explicitly by looking at bulk two- and three-point functions and their limit as one takes the insertion points to $\scri$. More generally, starting from a given $n$-point correlation function of massless bulk fields with spin indices suppressed for convenience,
\begin{equation}
\label{n-point correlator}
\langle \phi_1(X_1)\,...\, \phi_n(X_n) \rangle\,,
\end{equation}
we will take the limit in which all insertion points are sent out uniformly to $\scri^-$ and $\scri^+$,
\begin{equation}
\label{procedure}
\langle \phi^{\Delta,\pm}_1(x_1^\alpha)\,...\, \phi^{\Delta,\pm}_n(x_n^\alpha) \rangle\equiv \mathcal{N} \lim_{r \to \infty} r^{n \Delta}\, \langle \phi_1(\pm r,x_1^\alpha)\,...\, \phi_n(\pm r,x_n^\alpha) \rangle\,,
\end{equation}
where $\mathcal{N}$ is a normalisation of our choice. We expect this quantity to precisely take the form of a correlator for the corresponding Carrollian conformal fields. It should be emphasised that this is a priori different from existing proposals in the literature \cite{Banerjee:2018gce,Bagchi:2022emh,Donnay:2022wvx}, where scattering amplitudes map onto Carrollian conformal correlators after applying the integral transform \eqref{Fourier transform} with respect to the energy of the scattered states. We will come back to the relation between the two constructions in section~\ref{section 4}.  

As we have discussed in section~\ref{section 2}, Poincar\'e symmetry does not entirely fix the form of the Carrollian two- and three-point functions by contrast to standard conformal field theory. Similarly there is a variety of correlation functions with distinct causality structures that can be associated to a given set of relativistic bulk fields. On the one hand we can choose to send each bulk insertion point $X^\mu$ to either $\scri^+$ or $\scri^-$, a choice that we keep track of with the label~$\pm$  on the left-hand side of \eqref{procedure}. On the other hand the bulk correlator itself can be given various causality structures (Wightman, retarded/advanced or Feynman). We will observe that this freedom is precisely parametrised by the free coefficients of the Carrollian correlators \eqref{C2 solution}.

\subsection{Scalar two-point functions}
We start with the simple example of massless scalar two-point functions, which will however illustrate most of the interesting features of the correspondence between relativistic bulk correlators and Carrollian conformal correlators as defined through \eqref{procedure}. First we recall that the dual pair of fields in this case is simply
\begin{equation}
\phi(X^\mu) \quad \longleftrightarrow \quad \phi^\Delta(x^\alpha)\,,
\end{equation}
with the scaling dimension fixed to the value \eqref{Delta}. Indeed both types of fields encode the massless scalar particles.

We will apply the reduction formula \eqref{procedure} to the three common types of relativistic two-point functions known as Wightman, retarded, and Feynman propagators. We use the standard shorthand notation 
\begin{equation}
G(\lambda)\equiv \langle \phi(X_1)\phi(X_2)\rangle\,, \qquad \lambda \equiv (X_1-X_2)^2\,,
\end{equation}
even though the actual expression might also depends on the sign of $\delta t\equiv t_1-t_2$. Generalising the discussion in \cite{Dullemond:1984bc} to arbitrary spacetime dimension, the Wightman propagators are obtained from the homogeneous solution to the massless Klein--Gordon equation that vanishes at infinity, namely
\begin{equation}
\Psi(\lambda)=\frac{1}{\lambda^{\frac{d-1}{2}}}\,.
\end{equation}
Of course this is ill-defined on the lightcone $\lambda=0$. We can use time translation invariance $\delta t \to \delta t \pm i \epsilon$ to move the lightcone singularity off the physical spacetime domain, 
\begin{equation}
\label{Wightman}
G_\pm(\lambda)=\left(\frac{1}{\lambda \pm i\epsilon\, \sgn(\delta t)}\right)^{\frac{d-1}{2}}\,,
\end{equation}
thereby defining the two independent Wightman distributions. The retarded and Feynman Green's functions are then constructed using the standard field theory relations\begin{equation}
\label{G retared and Feynman}
\begin{split}
G_{\text{ret}}(\lambda)&=i\theta(\delta t)\left[G_+(\lambda)-G_-(\lambda) \right]=-2 \theta(\delta t)\, \text{Im}\, G_+(\lambda)\,,\\
i G_F(\lambda)&=\theta(\delta t)\, G_+(\lambda)+\theta(-\delta t)\, G_-(\lambda)\,.
\end{split}
\end{equation}
In particular the Feynman distribution always admits the simple analytic expression
\begin{equation}
iG_F(\lambda)=\left(\frac{1}{\lambda + i\epsilon}\right)^{\frac{d-1}{2}}\,.
\end{equation}

We now wish the to evaluate these propagators in the limit where the two insertion points are taken to $\scri$. For this it is useful to write the invariant $\lambda$ as well as the time interval $\delta t$ in retarded coordinates using \eqref{coord transf retarded},
\begin{equation}
\label{lambda}
\lambda \equiv (X_1-X_2)^2=2r_1r_2 |x_{12}|^2-2u_{12} (r_1-r_2)\,,
\end{equation}
\begin{equation}
\label{delta t}
\delta t\equiv t_1-t_2=\frac{1}{\sqrt{2}}\left(u_{12}+r_1(1+x_1^2)-r_2(1+x_2^2) \right)\,,
\end{equation}
where $u_{12}\equiv u_1-u_2$.
However the precise structure of the above distributions strongly depends upon whether $d$ is odd or even. Let us explicit study this for the Wightman propagators, from which the other propagators are constructed. For odd $d$ and thus integer $n=\frac{d-1}{2} \in \mathbb{N}$ the distributional meaning of \eqref{Wightman} is \cite{Kanwal}
\begin{equation}
\label{Gpm odd}
G_\pm(\lambda)=
\text{Pf}\left(\frac{1}{\lambda^n} \right)\pm i\pi (-1)^n\, \sgn(\delta t)\,  \delta^{(n-1)}(\lambda)\,, 
\end{equation}
in terms of the derivative of the Dirac distribution,
\begin{equation}
\delta^{(n)}(\lambda)\equiv \frac{d^n}{d\lambda^n} \delta(\lambda)\,,
\end{equation}
together with the pseudo-function $\text{Pf}\,(1/\lambda^n)$ that can be defined through the distributional derivative 
\begin{equation}
\label{pseudo function}
\langle\, \text{Pf}\left(\frac{1}{\lambda^{n+1}} \right),f(\lambda)\, \rangle=\frac{(-1)^n}{n!}\langle\,\frac{d^n}{d\lambda^n}\text{Pf}\left(\frac{1}{\lambda} \right),f(\lambda)\, \rangle=\frac{1}{n!} \langle\, \text{Pf}\left(\frac{1}{\lambda} \right),f^{(n)}(\lambda)\, \rangle\,,
\end{equation}
together with Cauchy's principal value distribution 
\begin{equation}
\text{Pf}\left(\frac{1}{\lambda}\right)=\frac{\lambda}{\lambda^2+0^+}\,.
\end{equation}
Roughly speaking $\text{Pf}\,(1/\lambda^n)$ smoothly interpolates between $1/\lambda^n$ for $\lambda \neq 0$ and zero at $\lambda=0$. For even $d$ on the other hand the distributional meaning of \eqref{Wightman} is rather \cite{GelfandShilov}
\begin{equation}
\label{Gpm even}
G_\pm(\lambda)=\frac{1}{|\lambda|^\frac{d-1}{2}}\left[\theta(\lambda)\pm i\, \sgn(\delta t)\, \theta(-\lambda)\right]\,.
\end{equation}
We use the above expressions in appendix~\ref{appendix} to recover in a very straightforward way the formula for the retarded Green's function available in the literature \cite{Hassani} for arbitrary spacetime dimension. In particular it is shown that the retarded Green's function is entirely supported on the lightcone for odd $d$ while it is supported within the lightcone for even $d$. See \cite{Balazs,Soodak1993WakesAW} and references therein for discussions and early derivations of this violation of Huygens principle. In what follows we take the insertion points to $\scri$, looking at the cases of even and odd spacetime dimensions separately.

\paragraph{Even spacetime dimensions.}
For odd $d$ the various bulk propagators are built out of linear combinations of the two basic distributions
\begin{equation}
\text{Pf}\left(\frac{1}{\lambda^n} \right) \qquad \text{and} \qquad \delta^{(n-1)}(\lambda)\,, 
\end{equation}
where $n=\frac{d-1}{2} \in \mathbb{N}$.
We are going to change variables $\lambda \mapsto |x_{12}|^2$ using \eqref{lambda} and evaluate these distributions in the limit $r\equiv r_1=r_2\to \infty$. This will allow us to evaluate the \textit{out-out} Carrollian correlators
\begin{equation}
\label{out out procedure}
\langle \phi^{\Delta,+}(x_1^\alpha)\, \phi^{\Delta,+}(x_2^\alpha) \rangle_{\pm/\text{ret}/F}\equiv \lim_{r_{1,2} \to \infty} (2r_1 r_2)^{\Delta}\, G_{\pm/\text{ret}/F}\,.
\end{equation}
Since in this case we have $\lambda=2r^2|x_{12}|^2$, for the first distribution we straightforwardly obtain
\begin{align}
\label{out out limit 1/lambda}
\lim_{r \to \infty} (2r^2)^n\, \text{Pf}\left(\frac{1}{\lambda^n} \right)= \text{Pf}\left(\frac{1}{|x_{12}|^{2n}} \right) =\frac{1}{|x_{12}|^{2n}}\,,
\end{align}
where the last expression is a shorthand notation.
For the derivative of the delta distribution we first explicitly perform the change of variable,
\begin{equation}
\delta^{(n-1)}(\lambda)=\left(\frac{d}{d\lambda}\right)^{n-1} \delta(\lambda)=\frac{1}{(2r^2)^n}\, \left(\frac{d}{dR}\right)^{n-1} \delta(R)\,,
\end{equation}
where $R=|x_{12}|^2$, such that using \eqref{delta t} we find
\begin{equation}
\lim_{r \to \infty} (2r^2)^n\, \sgn(\delta t)\, \delta^{(n-1)}(\lambda)= \sgn(u_{12})\, \delta^{(n-1)}(|x_{12}|^2)\,.
\end{equation}
Using these basic relations together with the expressions \eqref{Gpm odd} and \eqref{G retared and Feynman} for the bulk propagators, we can now explicitly evaluate \eqref{out out procedure} and obtain
\begin{equation}
\label{out out solutions}
\begin{split}
\langle \phi^{\Delta,+}(x_1^\alpha)\, \phi^{\Delta,+}(x_2^\alpha) \rangle_\pm&=\frac{1}{|x_{12}|^{2n}}\pm i\pi (-1)^n\, \sgn(u_{12})\,  \delta^{(n-1)}(|x_{12}|^2)\,,\\
\langle \phi^{\Delta,+}(x_1^\alpha)\, \phi^{\Delta,+}(x_2^\alpha) \rangle_{\text{ret}}&=2\pi (-1)^{n-1}\, \theta(u_{12})\,\delta^{(n-1)}(|x_{12}|^2)\,,\\
i\langle \phi^{\Delta,+}(x_1^\alpha)\, \phi^{\Delta,+}(x_2^\alpha) \rangle_F&=\frac{1}{|x_{12}|^{2n}} + i\pi (-1)^n\,  \delta^{(n-1)}(|x_{12}|^2)\,.
\end{split}
\end{equation}
We are but one step away from recovering the expressions \eqref{C2 solution bis} appropriate for a Carrollian conformal field $\phi^{\Delta,+}$ with scaling dimension $\Delta=n=\frac{d-1}{2}$.
To complete this step we just need to show that $\delta^{(n-1)}(|x|^2)$ reduces to the standard Dirac distribution $\delta(x^i)$ in $\mathbb{R}^{d-1}$. Working with an arbitrary integrable test function $\mathcal{F}(x)$ and introducing the variable $R\equiv |x|^2$, we have
\begin{equation}
\label{proof delta}
\begin{split}
&\int d^{d-1}x\, \mathcal{F}(x)\, \delta^{(n-1)}(|x|^2)=\frac{1}{2} \int_{S^{d-2}} d\hat x \int dR\, R^{n-1} \mathcal{F}(x)\,  \delta^{(n-1)}(R)\\
&=\frac{(-1)^{n-1}}{2}\int_{S^{d-2}} d\hat x \int dR\, \frac{d^{n-1}}{dR^{n-1}} \left(R^{n-1} \mathcal{F}(x)\right)  \delta(R)\\
&=\frac{(n-1)!(-1)^{n-1}\text{Vol}(S^{d-2})}{2}\, \mathcal{F}(0)\,.
\end{split}
\end{equation}
Thus we proved the distributional identity
\begin{equation}
\delta^{(n-1)}(|x|^2)=\frac{(n-1)!(-1)^{n-1}\text{Vol}(S^{d-2})}{2} \,\delta(x^i)\,,
\end{equation}
such that the correlators \eqref{out out solutions} are indeed of the form \eqref{C2 solution} as appropriate for a Carrollian conformal field $\phi^{\Delta,+}$ with scaling dimension $\Delta=n=\frac{d-1}{2}$. Moreover all three allowed solutions at this special value of the scaling dimension do actually appear. The \textit{in-out} correlators can be obtained in a similar fashion, with some additional complication however. Since in this case we set $r\equiv r_1=-r_2$, the invariant square distance \eqref{lambda} reduces to
\begin{equation}
\lambda=-2r^2|x_{12}|^2-4r u_{12}\,.
\end{equation}
Introducing the variable $\rho^2=r/(2 u_{12})$, we therefore have
\begin{equation}
(-2r^2)^n\, \text{Pf}\left(\frac{1}{\lambda^n} \right)= \frac{\rho^{2n}}{(1+\rho^2|x_{12}|^2)^n}\,.
\end{equation}
We recognize the above quantity as the bulk-boundary propagator of a scalar field in $AdS_d$. The limit $\rho \to \infty$ of this distribution is explicitly given by \cite{Donnay:2022ijr}
\begin{equation}
\frac{\rho^{2n}}{(1+\rho^2|x_{12}|^2)^n}=\frac{\pi^n}{\Gamma[n]} \left(\ln \rho^2-\psi(n)-\gamma_E\right) \delta(x_{12}^i)+\frac{1}{|x_{12}|^{2n}}+...\,,
\end{equation}
where $\psi(n)$ is the digamma function and $\gamma_E$ the Euler--Mascheroni constant. The above expression contains the expected $\ln \rho$ contact term \cite{Freedman:1998tz,Klebanov:1999tb} for this particular value of $n=\frac{d-1}{2}$ that precisely corresponds to the Breitenlohner--Freedman bound \cite{Breitenlohner:1982jf,Breitenlohner:1982bm}. Thus we find
\begin{equation}
\label{in out limit 1/lambda}
\lim_{r \to \infty} (-2r^2)^n\, \text{Pf}\left(\frac{1}{\lambda^n} \right)=  \frac{\pi^n}{\Gamma[n]} \left(\ln \frac{r}{2}-\ln u_{12}-\psi(n)-\gamma_E\right) \delta(x_{12}^i)+\frac{1}{|x_{12}|^{2n}}\,,
\end{equation}
and the Carrollian correlators take the form
\begin{equation}
\begin{split}
\langle \phi^{\Delta,+}(x_1^\alpha)\, \phi^{\Delta,-}(x_2^\alpha) \rangle_\pm&=\frac{\pi^n}{\Gamma[n]} \left(\ln \frac{r}{2}-\ln u_{12}-\psi(n)-\gamma_E\right) \delta(x_{12}^i)+\frac{1}{|x_{12}|^{2n}}\\
&\pm i\pi (-1)^{n-1}\,  \delta^{(n-1)}(|x_{12}|^2)\,,\\
\langle \phi^{\Delta,+}(x_1^\alpha)\, \phi^{\Delta,-}(x_2^\alpha) \rangle_{\text{ret}}&=2\pi (-1)^{n}\, \delta^{(n-1)}(|x_{12}|^2)\,,\\
i\langle \phi^{\Delta,+}(x_1^\alpha)\, \phi^{\Delta,-}(x_2^\alpha) \rangle_F&=\frac{\pi^n}{\Gamma[n]} \left(\ln \frac{r}{2}-\ln u_{12}-\psi(n)-\gamma_E\right) \delta(x_{12}^i)+\frac{1}{|x_{12}|^{2n}}\\
&+ i\pi (-1)^{n-1}\,  \delta^{(n-1)}(|x_{12}|^2)\,.
\end{split}
\end{equation}
Here no explicit $u$-dependence appears in the terms proportional $\delta^{(n-1)}(|x_{12}|^2)$ due to the definite time-ordering between in and out insertions. 

The contact term in the Wightman and Feynman correlators is problematic for two reasons. On the one hand it is explicitly divergent due to the $\ln r$ term, while on the other hand it does not satisfy the Carrollian Ward identities due to the $\ln u_{12}$ term. These two problems can be fixed at once by considering correlators involving at least one descendant field $\partial_u \phi^\Delta$ with scaling dimension $\Delta+1$ as described in \eqref{descendant}. Indeed in that case we find
\begin{equation}
\begin{split}
\langle \phi^{\Delta}(x_1^\alpha)\, \partial_u \phi^{\Delta,-}(x_2^\alpha) \rangle_\pm&=i\langle  \phi^{\Delta,+}(x_1^\alpha)\, \partial_u\phi^{\Delta,-}(x_2^\alpha) \rangle_F=\frac{\pi^n}{\Gamma[n]}\frac{\delta(x_{12}^i)}{u_{12}}\,,\\
\langle \phi^{\Delta,+}(x_1^\alpha)\, \partial_u \phi^{\Delta,-}(x_2^\alpha) \rangle_{\text{ret}}&=0\,,
\end{split}
\end{equation}
which are genuine Carrollian correlators of the form \eqref{g(u) solutions}. Of course we are allowed to take more derivatives $\partial_u$ if we wish. This should remind us of the massless boson in two dimensions which requires to consider descendant fields in order to avoid logarithmically divergent correlators. Another reason why it is physically reasonable to discard $\phi^\Delta$ in favour of $\partial_u \phi^\Delta$ is the Kirchhoff-d'Adh\'emar formula that allows to reconstruct the bulk field $\phi(X^\mu)$ from $\partial_u \phi^\Delta(x^\alpha)$ alone \cite{Donnay:2022wvx}.

\paragraph{Odd spacetime dimensions.}
For even $d$ all two-point functions are obtained from products of the two basic distributions
\begin{equation}
\frac{1}{|\lambda|^{\frac{d-1}{2}}} \qquad \text{and} \qquad \theta(\pm \lambda)\,.
\end{equation}
The computations of the Carrollian correlators as defined by \eqref{procedure} follow similar lines as in the previous section so we will be succint about it. 
For the \textit{out-out} insertions we observe that $\lambda\geq 0$ in the limit $r=r_1=r_2 \to \infty$, and using again \eqref{out out limit 1/lambda} we obtain
\begin{equation}
\begin{split}
\langle \phi^{\Delta,+}(x_1^\alpha)\, \phi^{\Delta,+}(x_2^\alpha) \rangle_\pm&=\frac{1}{|x_{12}|^{d-1}}=i\langle \phi^{\Delta,+}(x_1^\alpha)\, \phi^{\Delta,+}(x_2^\alpha) \rangle_F\,,\\
\langle \phi^{\Delta,+}(x_1^\alpha)\, \phi^{\Delta,+}(x_2^\alpha) \rangle_{\text{ret}}&=0\,,
\end{split}
\end{equation}
For \textit{in-out} insertions we have instead $\lambda \leq 0$, and using \eqref{in out limit 1/lambda} therefore yields
\begin{equation}
\langle \phi^{\Delta,+}(x_1^\alpha)\, \phi^{\Delta,-}(x_2^\alpha) \rangle_{\pm/\text{ret}/F}=\frac{\pi^{\frac{d-1}{2}}}{\Gamma[\frac{d-1}{2}]} \left(\ln \frac{r}{2}-\ln u_{12}-\psi(n)-\gamma_E\right) \delta(x_{12}^i)+\frac{1}{|x_{12}|^{d-1}}\,.
\end{equation}
In order to obtain genuine Carrollian correlators it is again more appropriate to consider at least one derivative $\partial_u$,
\begin{equation}
\label{odd in/out}
\langle \phi^{\Delta,+}(x_1^\alpha)\, \partial_u \phi^{\Delta,-}(x_2^\alpha) \rangle_{\pm/\text{ret}/F}=\frac{\pi^{\frac{d-1}{2}}}{\Gamma[\frac{d-1}{2}]} \frac{\delta(x_{12}^i)}{u_{12}}\,.
\end{equation}

\subsection{Photon and graviton two-point functions}
We now apply the same procedure to the photon and graviton propagators. Since they are constructed from the scalar propagators, they will inherit the features described in the previous subsection. We start from the propagators in momentum space in a generic gauge parametrised by $\xi$. For the photon propagator we use the standard formula
\begin{equation}
G_{\mu\nu}(p)=\left(\eta_{\mu\nu}-(1-\xi)\, \frac{p_\mu p_\nu}{p^2} \right) \frac{1}{p^2}\,,
\end{equation}
while the corresponding expression for the graviton propagator is given by \cite{Jakobsen:2020diz,Prinz:2020nru}
\begin{equation}
G_{\alpha\beta}^{\mu\nu}(p)=\left( P_{\alpha \beta}^{\mu\nu}-\frac{1}{d-1} \eta_{\alpha \beta}\, \eta^{\mu\nu}-2(1-\xi) P^{\mu\nu}_{\rho\kappa}\, \frac{p^\rho p_\sigma}{p^2} P^{\kappa\sigma}_{\alpha \beta}   \right)\frac{1}{p^2}\,,
\end{equation}
in terms of the projector onto the space of symmetric tensors
\begin{equation}
P^{\mu\nu}_{\alpha\beta}=\frac{1}{2}\left(\delta^\mu_\alpha \delta^\nu_\beta+\delta^\mu_\beta \delta^\nu_\alpha\right)\,.
\end{equation}
After Fourier transforming back to position space we obtain the bulk two-point functions
\begin{align}
\langle A_\mu(X_1) A_\nu(X_2) \rangle&=\left(\eta_{\mu\nu}-(1-\xi)\, \frac{X^{12}_\mu\, X^{12}_\nu}{(X_{12})^2}  \right) G(\lambda)\,,\\
\langle h^{\mu\nu}(X_1)\, h_{\alpha\beta}(X_2) \rangle&=\left( P_{\alpha \beta}^{\mu\nu}-\frac{1}{d-1} \eta_{\alpha \beta}\, \eta^{\mu\nu}-2(1-\xi) P^{\mu\nu}_{\rho\kappa}\, \frac{X_{12}^\rho\, X^{12}_\sigma}{(X_{12})^2} P^{\kappa\sigma}_{\alpha \beta}   \right) G(\lambda)\,,
\end{align}
where $G(\lambda)$ is any one of the scalar propagators given in \eqref{Wightman} and \eqref{G retared and Feynman}, depending on the choice of $i\epsilon$ prescription used to properly define these distributions. Using the identities 
\begin{equation}
\partial_i \hat q_1^\mu\, \eta_{\mu\nu}\, \partial_j \hat q_2^\nu=2 \delta_{ij}\,, \qquad X_1^\mu\, \eta_{\mu\nu}\, \partial_i \hat q^\nu_2=2 r_1 x^{12}_i\,,
\end{equation}
it is then straightforward to write down the two-point functions of the physical components \eqref{celestial components} in retarded coordinates, which simply yields
\begin{align}
\langle A_i(r,x_1^\alpha) A_j(r,x_2^\alpha) \rangle&=2\left(\delta_{ij}-(1-\xi)\, \frac{x^{12}_i\, x^{12}_j}{|x_{12}|^2}  \right) G_(\lambda)\,,\\
\langle h^{ij}(r,x_1^\alpha)\, h_{kl}(r,x_2^\alpha) \rangle&=4\left( P^{ij}_{kl}-2(1-\xi) P^{ij}_{mn}\, \frac{x_{12}^m\, x^{12}_p}{|x_{12}|^2} P^{np}_{kl}   \right) G(\lambda)\,.
\end{align}
At this point we already notice that the gauge choice $\xi=-1$ is such that the expressions in brackets reduce to the inversion tensors familiar from standard conformal field theory in $\mathbb{R}^{d-1}$. Indeed for $\xi=-1$ we have
\begin{align}
\langle A_i(r,x_1^\alpha) A_j(r,x_2^\alpha) \rangle&=2\,\mathcal{I}_{ij}(x_{12}) G(\lambda)\,,\\
\langle h_{ij}(r,x_1^\alpha)\, h_{kl}(r,x_2^\alpha) \rangle&=4\,\mathcal{I}_{im}(x_{12}) \mathcal{I}_{jn}(x_{12}) \mathcal{P}^{mn}_{kl}\, G(\lambda)\,,
\end{align}
with
\begin{equation}
\mathcal{I}_{ij}(x)=\delta_{ij}-\frac{2x_i x_j}{|x|^2}\,, \qquad \mathcal{P}^{mn}_{ij}=\frac{1}{2}\left(\delta^m_i \delta^n_j+\delta^m_j \delta^n_i\right)-\frac{1}{d-1}\, \delta^{mn} \delta_{ij}\,.
\end{equation}
Finally we need to apply the asymptotic reduction \eqref{procedure} in order to obtain the corresponding Carrollian correlators, which yields
\begin{align}
\langle A^{\Delta,\pm}_i(x_1^\alpha) A^{\Delta,\pm}_j(x_2^\alpha) \rangle&=\mathcal{I}_{ij}(x_{12}) \langle \phi^{\Delta,\pm}(x_1^\alpha)\, \phi^{\Delta,\pm}(x_2^\alpha) \rangle\,,\\
\langle h^{\Delta,\pm}_{ij}(x_1^\alpha)\, h^{\Delta,\pm}_{kl}(x_2^\alpha) \rangle&=\,\mathcal{I}_{im}(x_{12}) \mathcal{I}_{jn}(x_{12}) \mathcal{P}^{mn}_{kl}\, \langle \phi^{\Delta,\pm}(x_1^\alpha)\, \phi^{\Delta,\pm}(x_2^\alpha) \rangle\,,
\end{align}
in terms of the scalar Carrollian correlators described in the previous subsection. Such scalar correlators generically contain contact terms, in which case the inversion tensor $\mathcal{I}_{ij}$ reduces to the identity,
\begin{equation}
\mathcal{I}_{ij}(x)\, \delta^{(n-1)}(|x|^2)=\delta_{ij}\, \delta^{(n-1)}(|x|^2)\,, 
\end{equation}
using again the vanishing of the distribution $1/|x|^2\equiv \text{Pf}(1/|x|^2)$ at $|x|=0$. Similarly $\mathcal{I}_{im}(x) \mathcal{I}_{jn}(x) \mathcal{P}^{mn}_{kl}$ reduces to $\mathcal{P}_{ij,kl}$ which is the identity operator within the subspace of traceless symmetric $SO(d-1)$ tensors (generically called $\delta_{\sigma_1 \sigma_2}$ in section~\ref{section 2}).  

\subsection{An example of scalar three-point function}
In this section we consider one very specific example of bulk three-point correlator, and evaluate it in the limit where the insertion points are sent out to $\scri$. This is will provide us with an explicit example of holographic Carrollian three-point function. At tree-level such a three-point function is simply constructed by gluing together three propagators and amounts to considering the cubic coupling $\lambda\, \phi^3$ in the Lagrangian formalism. In momentum space the gluing corresponds to imposing momentum conservation, while in position space it amounts to integrate over the intermediate interaction point. For a massless scalar field this three-point function explicit reads
\begin{equation}
\langle \phi(X_1)\phi(X_2)\phi(X_3) \rangle= \int d^{d+1} X_4\, (\lambda_{14})^{-\frac{d-1}{2}} (\lambda_{24})^{-\frac{d-1}{2}} (\lambda_{34})^{-\frac{d-1}{2}}\,,
\end{equation}
with $\lambda_{ij}\equiv (X_i-X_j)^2$. For generic dimension $d$ an explicit expression for this integral is not available to our knowledge. However precisely at the value $d=5$ for which the cubic coupling $\lambda\, \phi^3$ is marginal and the theory conformal at tree-level, one can apply the star-triangle relation \cite{DEramo:1971hnd,Symanzik:1972wj}, in this case given by
\begin{equation}
\int d^6 X_4\, \frac{1}{(\lambda_{14})^2\, (\lambda_{24})^2\,(\lambda_{34})^2}
= \frac{\pi^3}{\lambda_{12}\, \lambda_{13}\,\lambda_{23}}\,.
\end{equation}
Strictly speaking this integral is only well-defined in euclidean signature. Analytic continuation to lorentzian signature with $\lambda_{ij} \mapsto \lambda_{ij}+i\epsilon$ then yields the time-ordered correlator
\begin{equation}
\label{Feynman 3-point}
\langle \phi(X_1)\phi(X_2)\phi(X_3) \rangle_{\mathcal{T}}=\frac{\pi^3}{(\lambda_{12}+i\epsilon)\, (\lambda_{13}+i\epsilon)\,(\lambda_{23}+i\epsilon)}\,.
\end{equation}
Using similar manipulations as in the previous subsections, it is not difficult to evaluate this expression in the limit where the insertion points are sent out to $\scri^+$. This yields
\begin{align}
\label{Carroll 3-point}
\langle \phi^{\Delta,+}(x_1^\alpha)\phi^{\Delta,+}(x_2^\alpha)\phi^{\Delta,+}(x_3^\alpha) \rangle_{\mathcal{T}}=\prod_{ij} \left(\frac{1}{|x_{ij}|^2}-i\pi\, \delta(|x_{ij}|^2) \right)=\frac{1}{|x_{12}|^2\, |x_{13}|^2 |x_{23}|^2}\,.
\end{align}
The second equality comes from the triviality of $\delta(|x|^2)$ in dimension $d=5$ unless it is multiplied by $1/|x|^2$. Indeed similar manipulations to \eqref{proof delta} yield
\begin{equation}
\int d^4x\, \mathcal{F}(x)\, \delta(|x|^2)=\frac{1}{2} \int_{S^{3}} d\hat x \int dR\, R\, \mathcal{F}(x)\, \delta(R)=0\,.
\end{equation}
Obviously the final expression in \eqref{Carroll 3-point} is a Carrollian three-point function of the form \eqref{standard 3-point} with $a=b=c=2=\Delta$. 

The case with one in and two out insertions is more involved but also more interesting, which corresponds to the correlator
\begin{equation}
\langle \phi^{\Delta,+}(x_1^\alpha)\phi^{\Delta,+}(x_2^\alpha)\phi^{\Delta,-}(x_3^\alpha) \rangle_{\mathcal{T}}\,.
\end{equation}
The factor in \eqref{Feynman 3-point} involving the two out insertions still simply evaluates to 
\begin{equation}
\frac{2r^2}{\lambda_{12}+i\epsilon}=\frac{1}{|x_{12}|^2}-i\pi \delta(|x_{12}|^2)\,.
\end{equation}
The remaining factor on the other hand gives 
\begin{equation}
\frac{(2r^2)^2}{(\lambda_{13}+i\epsilon)(\lambda_{23}+i\epsilon)}=\left( \frac{\rho_{13}^{2}}{1+\rho_{13}^2|x_{13}|^2} -i\pi \delta(|x_{13}|^2) \right)\left( \frac{\rho_{23}^{2}}{1+\rho_{23}^2|x_{23}|^2} -i\pi \delta(|x_{23}|^2) \right)\,,
\end{equation}
where we have again introduced $\rho^2_{ij}=r/(2u_{ij})$. As previously argued, a Dirac distribution $\delta(|x|^2)$ is trivial unless it is multiplied by $1/|x|^2$. Putting these elements together we thus have the following nontrivial distribution,
\begin{equation}
\frac{(2r^2)^3}{(\lambda_{12}+i\epsilon)\, (\lambda_{13}+i\epsilon)\,(\lambda_{23}+i\epsilon)}=\frac{1}{|x_{12}|^2} \frac{\rho_{13}^{2}}{1+\rho_{13}^2|x_{13}|^2}\frac{\rho_{23}^{2}}{1+\rho_{23}^2|x_{23}|^2}\,.
\end{equation}
In appendix~\ref{appendix B} we analyse in detail the asymptotic limit $r \to \infty$ of this distribution. As long as $|x_{12}|^2\gg r^{-1}$ it is simply given by
\begin{equation}
\langle \phi^{\Delta,+}(x_1^\alpha)\phi^{\Delta,+}(x_2^\alpha)\phi^{\Delta,-}(x_3^\alpha) \rangle_{\mathcal{T}}=\frac{1}{|x_{12}|^2|x_{13}|^2|x_{23}|^2}\,.
\end{equation}
However in the colinear limit $|x_{12}|^2 \ll r^{-1}$ we find instead
\begin{equation}
\begin{split}
\langle \phi^{\Delta,+}(x_1^\alpha)&\phi^{\Delta,+}(x_2^\alpha)\phi^{\Delta,-}(x_3^\alpha) \rangle_{\mathcal{T}}\\
&=\pi^2 \left(\ln \frac{r}{2}-\frac{1}{2} \ln (u_{13}\, u_{23}) -1 \right)\frac{\delta(x_{13}^i)}{|x_{12}|^2}+\frac{1}{|x_{12}|^2|x_{13}|^4}\,, \qquad (x_1 \sim x_2)\,.   
\end{split}
\end{equation}
This is explicitly divergent as in the case of the in-out two-point functions. To regulate it is enough to take one $\partial_u$ derivative, for example
\begin{equation}
\label{colinear}
\langle  \phi^{\Delta,+}(x_1^\alpha)\phi^{\Delta,+}(x_2^\alpha) \partial_u\phi^{\Delta,-}(x_3^\alpha) \rangle_{\mathcal{T}}=\frac{\pi^2}{2} \left(\frac{1}{u_{13}}+\frac{1}{u_{23}} \right)\frac{\delta(x_{13}^i)}{|x_{12}|^2}\,, \qquad (x_1 \sim x_2)\,.   
\end{equation}
As such this is a vanishing distribution since it has support on a vanishingly small region centered around $x_1\sim x_2$. This is actually closely related to the fact that the S-matrix element for three massless particles is also vanishing except in the colinear limit as we will discuss in section~\ref{section 4}. In order to get something nontrivial we can enhance the contribution \eqref{colinear} by multiplying it with a Dirac distribution $\delta(|x_{12}|^2)$. Since the latter has scaling dimension $2$, we need to compensate this by also multiplying it with a function such as $u_{13}\, u_{23}$ (there are many other possibilities), giving for example
\begin{equation}
\label{enhanced 3-point}
\langle  \phi^{\Delta,+}(x_1^\alpha)\phi^{\Delta,+}(x_2^\alpha) \partial_u\phi^{\Delta,-}(x_3^\alpha) \rangle_{\mathcal{T}}^{\text{enhanced}}=\frac{\pi^2}{2} \left(u_{13}+u_{23} \right)\delta(x_{13}^i)\delta(x_{12}^i)\,.   
\end{equation}
This prescription yields a Carrollian conformal correlator of the form \eqref{two contact}. In summary, for in-out correlators we again had to consider at least one descendant field $\partial_u \phi^\Delta$ in place of $\phi^\Delta$ in order to get rid of anomalous $\ln r$ divergences. In that case we find a nonzero expression only in the colinear regime, just as it happens for the corresponding S-matrix element. As such this correlator is therefore a vanishing distribution, but if we insist on extracting a Carrollian conformal three-point function we can enhance the colinear contribution by multiplying it with a (scale-invariant) Dirac distribution.  

\section{Massless scattering amplitudes}
\label{section 4}
In this paper we have classified  the two- and three-point functions that Carrollian conformal fields encoding the massless particle UIRs of $ISO(1,d)$ can have. We have also observed that such correlators are obtained by evaluating correlators of the corresponding massless relativistic fields in $\mathbb{M}^{d+1}$ in the limit where the insertion points are sent out to $\scri$. In this section to scattering amplitudes which constituted one of the main motivations for carrying out this work. 

As previously mentioned a proposal has been put forward to relate scattering amplitudes and Carrollian conformal correlators \cite{Banerjee:2018gce,Bagchi:2022emh}, which simply consists in applying the Fourier-like transform \eqref{Fourier transform} to the S-matrix elements themselves,
\begin{equation}
\label{proposal}
\langle \phi_1^{\Delta_1,+}(x_1^\alpha)\, ...\, \phi_n^{\Delta_n,-}(x_n^\alpha) \rangle \equiv \prod_{k=1}^n \int_0^\infty \dd \omega_k\, \omega_k^{\Delta_k-1} e^{-i \eta_k \omega_k u_k}\, \langle\, p_1(\omega_1,x_1^i)\,...|\mathcal{S}|...\, p_n(\omega_n,x_n^i)\, \rangle \,,
\end{equation}
where $\eta_k=\pm 1$ depending whether the corresponding particle is outgoing or ingoing. The simplest example to consider is obviously that of the free 1-1 scattering, with S-matrix element simply given by the Lorentz-invariant inner product between two onshell massless particles,
\begin{equation}
\langle\, p_1|\mathcal{S}|\, p_2\, \rangle=|\vec p_1|\, \delta(\vec p_1- \vec p_2)=\omega_1^{2-d}\, \delta(\omega_1-\omega_2)\, \delta(x_1^i-x_2^i)\,.
\end{equation}
The second expression follows from the parametrisation of massless momenta given in \eqref{momentum parametrisation}.
In that case the expression \eqref{proposal} reduces to
\begin{equation}
\label{divergent integral}
\begin{split}
\langle \phi^{\Delta_1,+}(x_1^\alpha) \phi^{\Delta_2,-}(x_2^\alpha) \rangle&=\int_0^\infty \dd \omega_1 \dd \omega_2\, \omega_1^{\Delta_1-1} \omega_2^{\Delta_2-1} e^{-i\omega_1 u_1}e^{i\omega_2 u_2}\, \langle\, p_1|\mathcal{S}|\, p_2\, \rangle\\
&=\delta(x_{12}^i) \int_0^\infty \dd \omega\,  \omega^{\Delta_1+\Delta_2-d}\, e^{-i\omega u_{12}}\,.
\end{split}
\end{equation}
For convergence of the above integral, we need to perform the analytic continuation $u_{12} \mapsto u_{12}-i \epsilon$, and we obtain \cite{Banerjee:2018gce,Liu:2022mne,Bagchi:2022emh,Donnay:2022wvx}
\begin{equation}
\label{free transformed amplitude}
\langle \phi^{\Delta_1,+}(x_1^\alpha) \phi^{\Delta_2,-}(x_2^\alpha) \rangle=\Gamma[\Delta_1+\Delta_2+1-d]\, \frac{\delta(x_{12}^i)}{(iu_{12})^{\Delta_1+\Delta_2+1-d}}\,.
\end{equation}
We observe that this indeed takes the form of a Carrollian conformal correlator \eqref{g(u) solutions} provided $\Delta_1+\Delta_2 \neq d-1$. For the special value $\Delta_1=\Delta_2 = \frac{d-1}{2}$ which precisely arises from the holographic correspondence with massless bulk fields, the integral \eqref{divergent integral} is still not convergent, which is also seen in \eqref{free transformed amplitude} from the corresponding pole in $\Gamma[\Delta_1+\Delta_2+1-d]$.  The amplitude \eqref{free transformed amplitude} can be expanded in $\beta\equiv \Delta_1+\Delta_2+1-d$ for small $\beta$, and diverges as $\beta^{-1}$ in the strict $\beta \to 0$ limit\cite{Liu:2022mne,Donnay:2022wvx},
\begin{equation}
\langle \phi^{\Delta,+}(x_1^\alpha) \phi^{\Delta,-}(x_2^\alpha) \rangle=\left(\frac{1}{\beta}-\gamma_E-\ln (iu_{12})+O(\beta^2) \right) \delta(x_{12}^i)\,, \qquad \Delta=\frac{d-1}{2}\,.
\end{equation}
We observe that this expression contains the same kind of contact term divergence as in \eqref{in out limit 1/lambda} with $1/\beta$ playing the role of $\ln r$. In section~\ref{section 3} we have argued that we should consider descendant fields such that $\Delta_1+\Delta_2 >d-1$, in which case \eqref{free transformed amplitude} is regular and further agrees up to a choice of normalisation with the independently derived holographic Carrollian correlator \eqref{odd in/out}.

We turn now to the scattering of three massless particles, say one ingoing and two outgoing, with S-matrix element simply given by a momentum-conserving delta function. It is well-known that three onshell massless particles can only satisfy momentum conservation if they are colinear. Hence even as a tempered distribution the three-point amplitude vanishes unless we decide to enhance the contribution coming from this colinear limit, which is done in practice by multpliying the S-matrix element by an additional Lorentz-invariant Dirac distribution \cite{Bagchi:2023fbj}. If we align the two outgoing particles, the corresponding enhanced S-matrix element is then given by
\begin{equation}
\label{enhanced S matrix}
\langle\, p_1,p_2\, |\mathcal{S}|\, p_3\, \rangle^{\text{enhanced}}=\omega_3^{2-d}\, \delta(\omega_1+\omega_2-\omega_3)\, \delta(x_{13}^i)\, f(\omega_i,x_i)\,\delta(x_{12}^i)\,.
\end{equation}
The factor $\omega_3^{2-d}\, \delta(\omega_1+\omega_2-\omega_3)=|\vec p_3|\, \delta(\vec p_1+\vec p_2-\vec p_3)$ is just the momentum conserving delta function, while the remaining factor is enhancing the colinear contribution. The function $f(\omega_i,x_i)$ is chosen to preserve Lorentz invariance and must therefore have scaling dimension $1-d$ but is otherwise arbitrary. For illustration we choose here 
$f(\omega_i,x_i)=(\omega_1 \omega_2)^{\frac{1-d}{2}}$. Applying now the transform \eqref{proposal}, we get
\begin{equation}
\begin{split}
C^{(3)}&\equiv \langle \phi^{\Delta_1,+}(x_1^\alpha) \phi^{\Delta_2,+}(x_2^\alpha)\phi^{\Delta_3,-}(x_3^\alpha) \rangle\\
&=\delta(x_{12}^i)\delta(x^i_{13})\int_0^\infty \dd \omega_1 \dd \omega_2\, \omega_1^{\Delta_1-\frac{d+1}{2}} \omega_2^{\Delta_2-\frac{d+1}{2}} (\omega_1+\omega_2)^{\Delta_*}e^{-i\omega_1 u_{13}}e^{-i\omega_2 u_{23}}\\
&=\delta(x_{12}^i)\delta(x^i_{13})\sum_{k=0}^{\Delta_*} \binom{\Delta^*}{k} \int_0^\infty \dd \omega_1 \dd \omega_2\, \omega_1^{\Delta_1-\frac{d+1}{2}+k} \omega_2^{\Delta_2-\frac{d+1}{2}+\Delta_*-k} e^{-i\omega_1 u_{13}}e^{-i\omega_2 u_{23}}\,,
\end{split}
\end{equation}
where we have assumed $\Delta_*\equiv \Delta_3-d+1 \in \mathbb{N}_+$ in order to perform a binomial expansion of $(\omega_1+\omega_2)^{\Delta_*}$ as done in \cite{Bagchi:2023fbj}. Like before we ensure convergence of the remaining integrals by performing the analytic continuation $u_{13} \mapsto u_{13}-i\epsilon$ and $u_{23} \mapsto u_{23}-i\epsilon$, and we obtain 
\begin{equation}
C^{(3)}=\delta(x_{12}^i)\delta(x^i_{13})\sum_{k=0}^{\Delta_*} \binom{\Delta^*}{k} \frac{\Gamma[\Delta_1-\frac{d-1}{2}+k]\,\Gamma[\Delta_2-\frac{d-1}{2}+\Delta_*-k]}{(iu_{13})^{\Delta_1-\frac{d-1}{2}+k}(iu_{23})^{\Delta_2-\frac{d-1}{2}+\Delta_*-k}}\,.
\end{equation}
This is manifestly a sum of Carrollian conformal correlators of the form \eqref{two contact}. Here we do not find agreement with the enhanced holographic correlator \eqref{enhanced 3-point} simply because the enhancing prescription is not unique. We do have agreement at the level of the non-enhanced correlators though, namely that they have vanishingly small support corresponding to the colinear limit $x_1 \sim x_2$. 

In summary, we find that it is necessary to consider conformal correlators involving at least one descendant Carrollian field with $\Delta\neq \frac{d-1}{2}$. In this case we find agreement between two approaches \eqref{procedure} and \eqref{proposal} to construct two- and three-point Carrollian conformal correlators. The three-point function however vanishes as a distribution, which raises questions regarding the construction of a Carrollian OPE. In the standard approach to scattering amplitudes, this issue is circumvented by the fact that virtual cubic interactions happen offshell. One potential way around this issue within a strictly onshell formalism is to consider complex momenta, i.e., to analytically continue the celestial sphere $\mathbb{R}^{d-1}$ to $\mathbb{C}^{d-1}$ in order to give non-vanishing support to the three-point scattering amplitude and corresponding Carrollian correlator \cite{Salzer:2023jqv}. We hope to come back to this point in a future publication.

\section*{Acknowledgments}
I thank Jos\'e Figueroa--O'Farrill, Yorgo Pano, Charlotte Sleight, Massimo Taronna and especially Jakob Salzer for useful discussions. I also thank Giovanni Spinielli for spotting sign typos. This work is supported by a Postdoctoral
Research Fellowship granted by the F.R.S.-FNRS (Belgium).

\appendix
  
\section{Retarded Green's functions in Minkowski space}
\label{appendix}
In this short appendix we show that the formula for the massless retarded Green's functions in $\mathbb{M}^{d+1}$ available in the literature \cite{Hassani} can be easily recovered from the expressions \eqref{Wightman}-\eqref{G retared and Feynman} together with \eqref{Gpm odd} and \eqref{Gpm even}. For odd $d$ and $n=\frac{d-1}{2} \in \mathbb{N}$ we simply need to use the transformation rules for the delta distribution when changing variable from $\lambda$ to $r=|\vec x_1-\vec x_2|$,
\begin{equation}
\delta^{(n-1)}(\lambda)=\left(\frac{1}{2r}\frac{d}{dr} \right)^{n-1} \left[\frac{\delta(r-\delta t)+\delta(r+\delta t)}{2r}\right]\,.
\end{equation}
Using \eqref{G retared and Feynman} and \eqref{Gpm odd} we obtain for the retarded Green's function
\begin{equation}
\begin{split}
G_{\text{ret}}(\lambda)&=2\pi \left(-\frac{1}{2r}\frac{d}{dr} \right)^{n-1} \left[\frac{\delta(r-\delta t)}{2r}\right]\,.
\end{split}
\end{equation}
Up to normalisation this is in agreement with the textbook formula \cite{Hassani}, although the derivation given here is much simpler. For even $d$ and $n=d/2 \in \mathbb{N}$, using \eqref{G retared and Feynman} and \eqref{Gpm even} we can write
\begin{equation}
G_{\text{ret}}(\lambda)=-\frac{2\theta(\delta t)\theta(-\lambda)}{|\lambda|^{n-\frac{1}{2}}}\,,
\end{equation}
or equivalently
\begin{equation}
G_{\text{ret}}(\lambda)=-\frac{2\theta(\delta t)\theta(-\lambda)}{(\delta t^2-r^2)^{n-\frac{1}{2}}}=-\frac{2\theta(\delta t)\theta(-\lambda)}{n-1!!}\left(-\frac{1}{r} \frac{d}{dr} \right)^{n-1} \left[\frac{1}{\sqrt{\delta t^2-r^2}}\right]\,.
\end{equation}
Up to normalisation this is again in agreement with the textbook formula \cite{Hassani}.

\section{Asymptotics of the distribution $D_\beta$}
\label{appendix B}
In this appendix we want to analyse the quantity
\begin{equation}
D_\beta(x_3)=\left( \frac{\rho_{13}^{2}}{1+\rho_{13}^2|x_{13}|^2} \right)^{1+\beta}\left( \frac{\rho_{23}^{2}}{1+\rho_{23}^2|x_{23}|^2}\right)^{1+\beta}\,,
\end{equation}
regarded as a distribution over $x_3^i$. In particular we want to analyse its asymptotic behavior in the limit $\rho_{13}, \rho_{23} \to \infty$ and fixed $\chi=\rho_{13}^2/\rho_{23}^2$. The answer will vary depending whether $|x_{12}| \gg \rho^{-1}$ or $|x_{12}|\lesssim \rho^{-1}$ in that limit.

\paragraph{Non-colinear regime.}
We start by assuming that the external point $x_1$ and $x_2$ have nonvanishing separation $|x_{12}| \gg \rho^{-1}$. Then as long as $|x_{13}|^2\gg \rho_{13}^{-2}$ and $|x_{23}|^2\gg \rho_{23}^{-2}$, we simply have
\begin{equation}
\label{D generic}
D_\beta(x_3) \approx \frac{1}{|x_{13}|^{2+2\beta}|x_{23}|^{2+2\beta}}\,.
\end{equation}
However in the small region $\mathcal{D}_1$ centered around $x_1$ and characterised by $|x_{13}|^2 \lesssim \rho_{13}^{-2}$ and $|x_{23}|^2\gg \rho_{23}^{-2}$, there is no reason \eqref{D generic} should hold. In this region integration against a test function $\mathcal{F}(x_3)$ can be approximated as
\begin{equation}
\begin{split}
\int_{\mathcal{D}_1} d^4 x_3\, \mathcal{F}(x_3) D_\beta(x_3)&\approx \frac{\mathcal{F}(x_1)}{|x_{12}|^2} \int d^4 x_3 \left(\frac{\rho_{13}^{2}}{1+\rho_{13}^2|x_{13}|^2} \right)^{1+\beta}\\
&=\frac{\mathcal{F}(x_1)}{|x_{12}|^2}\frac{\pi^2\Gamma[\beta-1]}{\Gamma[\beta+1]} (\rho_{13})^{2(\beta-1)}\,,
\end{split}
\end{equation}
which is subleading as $\rho_{13} \to \infty$. Hence we do not get a nonzero contribution as leading order from the region $\mathcal{D}_1$. A similar story holds for the region $\mathcal{D}_2$ centered around $x_2$. 

\paragraph{Colinear regime.} We now analyse the case where the external points $x_1$ and $x_2$ have vanishing separations $|x_{12}|\ll \rho^{-1}$. Then as long as $|x_{13}|^2\gg \rho_{13}^{-2}$ and $|x_{23}|^2\gg \rho_{23}^{-2}$, we simply have
\begin{equation}
D_\beta(x_3) \approx \frac{1}{|x_{13}|^{2+2\beta}|x_{23}|^{2+2\beta}} \approx  \frac{1}{|x_{13}|^{4+4\beta}}\,.
\end{equation}
In the small region $\mathcal{D}$ centered around $x_1\sim x_2$ and characterised by $|x_{13}|^2 \ll \rho_{13}^{-2}$, integration against a test function $\mathcal{F}(x_3)$ can be approximated as
\begin{equation}
\begin{split}
\int_{\mathcal{D}} d^4 x_3\, \mathcal{F}(x_3) D_\beta(x_3)&\approx \mathcal{F}(x_1) \int d^4 x_3\, D_\beta(x_3)\,.
\end{split}
\end{equation}
The above integral admits the explicit series representation \cite{Leonhardt:2003qu}
\begin{equation}
\begin{split}
\int d^4 x_3\, &D_\beta(x_3)=\frac{\pi^2}{\Gamma[1+\beta]^2}\, (\rho_{13}\, \rho_{23})^{2\beta}\\
&\times \sum_{k \in \mathbb{N}} \frac{\sigma^k}{k!} \Big( \chi^{\beta}\, \frac{\Gamma[1-\beta]\Gamma[2\beta+k]\Gamma[1+\beta+k]}{\Gamma[2+k]}\,  {}_2F_1[2\beta+k,1+\beta+k,\beta;\chi]\\
&\qquad \qquad \, +\chi\, \Gamma[\beta-1]\Gamma[1+\beta+k]\, {}_2F_1[1+\beta+k,2+k,2-\beta;\chi] \Big)\,,
\end{split}
\end{equation}
written in terms of the variables $\sigma \equiv -\rho_{13}^2|x_{12}|^2$ and $\chi \equiv \rho_{13}^2/\rho_{23}^2$. In the regime we are interested in, $|\sigma| \ll 1$ such that the first term dominates the sum,
\begin{equation}
\begin{split}
\int d^4 x_3\, &D_\beta(x_3)\approx \frac{\pi^2}{\Gamma[1+\beta]}\, (\rho_{13}\, \rho_{23})^{2\beta}\\
&\times  \Big( \chi^{\beta}\, \Gamma[1-\beta]\Gamma[2\beta]\,  {}_2F_1[2\beta,1+\beta,\beta;\chi] +\chi\, \Gamma[\beta-1]\, {}_2F_1[1+\beta,2,2-\beta;\chi] \Big)\\
&= \frac{\pi^2}{2\beta(1+2\beta)}\, (\rho_{13}\, \rho_{23})^{2\beta} \chi^{-\beta}  {}_2F_1[2\beta,1+\beta,2+2\beta;1-\chi^{-1}]\\
&= \frac{\pi^2}{2\beta(1+2\beta)}\, (\rho_{13}\, \rho_{23})^{2\beta} \chi^\beta  {}_2F_1[2\beta,1+\beta,2+2\beta;1-\chi]\,,
\end{split}
\end{equation}
and the last equality shows that it is actually symmetrical under exchange of indices $1 \leftrightarrow 2$ as it should. Hence in the colinear limit $x_1 \sim x_2$, the asymptotic behavior of the distribution $D_\beta(x_3)$ is given by
\begin{equation}
\label{asymptotic Db}
D_\beta(x_3)\approx \frac{\pi^2}{2\beta(1+2\beta)}\, (\rho_{13}\, \rho_{23})^{2\beta} \chi^\beta  {}_2F_1[2\beta,1+\beta,2+2\beta;1-\chi]\, \delta(x^i_{13})+\frac{1}{|x_{13}|^{4+4\beta}}\,.
\end{equation}
In the main text we are interested in the $\beta \to 0$ limit of this distribution. To achieve this we need the small $\beta$ expansion of the quantities appearing in the above expression. The Taylor expansion of the hypergeometric function is easily evaluated using the first few identities given in \cite{Ancarani:2009zz},
\begin{equation}
\chi^\beta {}_2F_1[2\beta,1+\beta,2+2\beta;1-\chi]=1+O(\beta^2)\,,
\end{equation}
together with 
\begin{equation}
x^\beta=e^{\beta \ln x}=1+\beta \ln x+O(\beta^2)\,.
\end{equation}
We also have the distributional identity \cite{Donnay:2022ijr} 
\begin{equation}
\frac{1}{|x_{13}|^{4+4\beta}}=-\frac{\pi^2}{2\beta}\, \delta(x_{13}^i)+\frac{1}{|x_{13}|^4}+O(\beta^2)\,, \qquad (\beta <0)\,,
\end{equation}
such that the $1/\beta$ poles cancel out between the two terms in \eqref{asymptotic Db}, producing the finite expression
\begin{equation}
\lim_{\beta \to 0^-} D_\beta(x_3)\approx \pi^2 \left(\ln \rho_{13}+\ln \rho_{23} -1 \right)\delta(x^i_{13})+\frac{1}{|x_{13}|^4}\,,
\end{equation}
still understood as the leading contribution in the colinear limit $x_1 \sim x_2$ at large $\rho_{13},\rho_{23}$.

\bibliography{bibl}
\bibliographystyle{JHEP}  
\end{document}